\documentclass[9pt,twocolumn,twoside]{osajnl}

\journal{ol} 

\usepackage{tikz}

\setboolean{shortarticle}{true} 

\title{Modal analysis using photonic lanterns coupled to arrays of waveguides}
\author[1,*]{Momen Diab}
\author[2]{Stefano Minardi}

\affil[1]{innoFSPEC - Leibniz Institute for Astrophysics Potsdam, An der Sternwarte 16, 14482 Potsdam, Germany.}
\affil[2]{LASOS GmbH, Franz-L\"owen Str. 2, 07745 Jena, Germany.}

\affil[*]{Corresponding author: mdiab@aip.de}

\ociscodes{(120.5050) Phase measurement, (120.4640) Optical instruments, (030.4070) Modes, (080.1238) Array waveguide devices, (060.2420) Fibers.}

\begin{abstract}
We present a new concept of an integrated optics component capable of measuring the complex 
amplitudes of the modes at the tip of a multimode waveguide.
The device uses a photonic lantern to split the optical power carried by an $N$-modes waveguide among a
collection of single-mode wave\-guides that excite a periodic array of at least $N^2$ single-mode evanescently-coupled wave\-guides. The power detected at each output of the array 
is a linear combination of the products of the modal amplitudes\textemdash a relation that can, under suitable conditions, be inverted allowing the derivation of the amplitudes and relative phases of the 
modal mixture at the input. The expected performance of the device is discussed and its application to the 
real-time measurement of modal instability in high power fiber lasers is proposed. 

\end{abstract}

\setboolean{displaycopyright}{true}

\begin{document}

\maketitle


\definecolor{mycolor1}{RGB}{36,111,237}
 \newcommand{\blue}{\raisebox{2pt}{\tikz{\draw[-,black!00!mycolor1,solid,line width = 2.0pt](0,0) -- (5mm,0) ;}}}

  \newcommand{\red}{\raisebox{1pt}{\tikz{ \draw[-,red,solid,line width = 0.8pt](0.,0.0mm) -- (2.0mm,0.0mm); \draw[red,solid,line width = 0.8pt](2.0mm,-0.7mm) rectangle (3.4mm,0.7mm);\draw[-,red,solid,line width = 0.8pt](3.4mm,0.0mm) -- (5.4mm,0.0mm);}}}
  
  \newcommand{\blueline}{\raisebox{2pt}{\tikz{\draw[-,black!40!blue,solid,line width = 2.0pt](0,0) -- (1.5mm,0) ;}}}

\definecolor{mycolor3}{RGB}{68,161,68}
\newcommand{\green}{\raisebox{2pt}{\tikz{\draw[-,black!00!mycolor3,dashed,line width = 0.9pt](0,0) -- (5mm,0) ;}}}

\definecolor{mycolor4}{RGB}{120,171,48}
\newcommand{\greenbox}{\raisebox{0pt}{\tikz{\path[fill = black!00!mycolor4,opacity=1,line width = 1.2pt](2.mm,0) rectangle (7.0mm,2.5mm);\draw[black,solid,line width = 1.0pt](2.0mm,0) rectangle (7.0mm,2.5mm);}}}

\definecolor{mycolor5}{RGB}{127,184,222}
\newcommand{\bluebox}{\raisebox{0pt}{\tikz{\path[fill=black!00!mycolor5,opacity=1](2.mm,0) rectangle (7.0mm,2.5mm);\draw[black,dashed,line width = 0.5pt](2.0mm,0) rectangle (7.0mm,2.5mm);}}}

\definecolor{mycolor6}{RGB}{222,125,0}
\newcommand{\orange}{\raisebox{0pt}{\tikz{ \draw[-,mycolor6,solid,line width = 0.8pt](0.,0) -- (2.0mm,0); \draw[mycolor6,solid,line width = 0.8pt](2.7mm,0) circle (0.7mm);\draw[-,mycolor6,solid,line width = 0.8pt](3.4mm,0) -- (5.4mm,0);}}}

\definecolor{mycolor7}{RGB}{0,0,0}
\newcommand{\black}{\raisebox{0pt}{\tikz{ \draw[-,mycolor7,solid,line width = 0.8pt](0.,0) -- (2.0mm,0); \draw[mycolor7,solid,fill=black,line width = 0.8pt](2.6mm,0) circle (0.5mm);\draw[-,mycolor7,solid,line width = 0.8pt](3.2mm,0) -- (5.2mm,0);}}}

\definecolor{mycolor2}{RGB}{126,47,142}
\newcommand{\purple}{\raisebox{2pt}{\tikz{ \draw[-,mycolor2,solid,line width = 0.8pt](0,0) -- (2.0mm,0); \draw[mycolor2,solid,line width = 0.8pt](2.7mm,0) circle (0.1mm);\draw[-,mycolor2,solid,line width = 0.8pt](3.4mm,0) -- (5.4mm,0);}}}

\newcommand\solidrule[1][0.5cm]{\rule[0.5ex]{#1}{1.0pt}}
\newcommand\dashedrule{\mbox{%
  \solidrule[1mm]\hspace{1mm}\solidrule[1mm]\hspace{1mm}\solidrule[1mm]\hspace{1mm}}}


The decomposition of the optical field carried by a multimode fiber into its supported modes is a requirement for several applications in the photonic industry ranging from 
spatial division multiplexing communication protocols \cite{Li:2017} to management of modal instability in high power fiber lasers \cite{Eidam:2011}.  
However, while modal decomposition of an optical field is a straightforward numerical task, it is usually a rather complex one from the experimental viewpoint.  
Free space modal characterization techniques often use 
phase holograms as multiplexed mode matched filters \cite{Goodman:1968} 
to retrieve amplitude and phase from optical cross-correlations of the mode functions with the fiber field \cite{Kaiser:2009}.
Direct field measurement by means of a wavefront sensor \cite{Paurisse:2012} or by means of a spatial-spectral scan of the mode profile \cite{Nicholson:2008} are other free space methods used to 
estimate the modal coefficients in an excited fiber. 
A common feature of free space methods is 
the requirement of extended setups and slow data processing, 
making them unsuitable for high speed applications.
Integrated optics mode-division multiplexer/demultiplexer can offer devices with smaller footprints and a fast measurement of the modulus of the modal (complex) amplitude.
In this respect, asymmetric couplers in 2 \cite{Narevicius:2005} and 3 dimensions \cite{Riesen:2014,Hanzawa:2014} and asymmetric photonic lanterns \cite{LeonSaval:2014} have been used to 
multiplex/demultiplex up to 3 modes in rectangular multimode waveguides. While the speed of the modal decomposition in 
demultiplexers is limited by the speed of the light detector at the output, an integrated device measuring the relative phase between the excited modes has not been reported so far.
The measurement of the modal phase is an essential requirement to determine the exact optical pattern at the output plane of the multimode fiber, information that could find application 
in the active control of modal instability in high power fiber lasers \cite{Otto:2013}.



\begin{figure}[h!]
\centering
\includegraphics[width=0.9\linewidth]{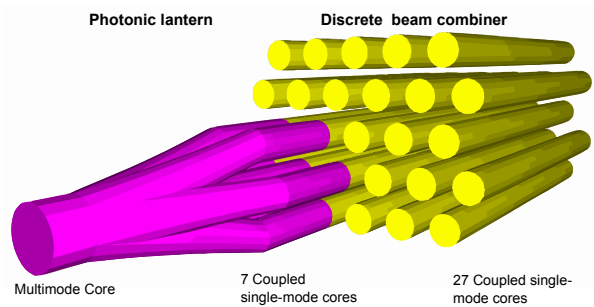}
\caption{Layout of a photonic mode analyzer designed to retrieve the complex amplitudes of the modes excited at its multimode input. The scale in the propagation direction is reduced by 200x relative to the transverse coordinates and the cladding of the waveguides is hidden for clarity.}
\label{fig:layout1}
\end{figure}

In this letter we present the concept of a new integrated optics device designed to 
accomplish a complete measurement of the complex modal amplitudes 
at the input of a multimode wave\-guide \cite{patent}.
The proposed device combines a photonic lantern \cite{LeonSaval:2005} with a discrete beam combiner \cite{Minardi:2010}, as sketched in Fig. \ref{fig:layout1}.
After describing the operating principle, we present the results of a numerical 
optimization of the device and discuss estimates of the precision of the modal complex
amplitude retrieval relative to detection noise.     

As known, the photonic lantern is a tapered waveguide that transforms 
adiabatically the modes supported by its multimode end into the supermodes of an 
array of single-mode wave\-guides. 
The output of the lantern can be used to excite an array of coupled single-mode 
waveguides, designed to act as an interfero\-metric beam combiner, i.e., the discrete beam combiner. We show here below that by choosing appropriately the interface between the photonic lantern and the discrete beam combiner, it is possible to relate 
linearly the power carried by each output of the beam combiner with the square 
moduli and mutual products of the input modal amplitudes.   

We begin by describing our system in terms of a linear operator  
$\hat{\psi}$ from the $N$-dimensional functional space spanned by the transverse modes $U_i(x,y)$ of the multimode end of the lantern to an $M$-dimensional functional space describing the excitations of the $M$ waveguides of the discrete beam combiner. 
In the frame of the coupled-mode approximation, the fields supported by the array of evanescently-coupled waveguides are written in terms of the mode profile $u_0(x,y)$ of the isolated single-mode waveguide, each centered at the coordinates $(x_n,y_n)$ of the $n$th waveguide axis. 
With this approximation, we can choose as a base of the $M$-dimensional space, 
fields composed of single waveguide excitations at the output of the array 
$u_n(x,y) = u_0(x-x_n,y-y_n)$. We can thus rewrite the operator $\hat{\psi}$ as an $M\times N$ 
complex matrix $\psi_{n,i} = \langle u_n|\hat\psi U_i\rangle$, where the norm 
$\langle\cdot|\cdot\rangle$ is the usual overlap integral. Notice that $\psi_{n,i}$ represents the 
complex amplitude of the mode 
of the $n$th waveguide at the output of the array due to an excitation of the photonic 
lantern input with mode $U_i$. If the transition from multimode end to array output is lossless 
and the chosen bases are normalized we clearly have that $\sum_{n=1}^M|\psi_{n,i}|^2 = 1$ for every $i = 1, ..., N$. 
A generic superposition of the input modes $U_i$ with weights $c_i\in\mathbb{C}$ will
therefore map into the output amplitude 
\begin{align}
\phi_n = \sum_{i=1}^{N} c_i\psi_{n,i}, && n=1,...,M.
\label{eq:refname1}
\end{align}
We now consider that powers rather than fields are measured at the output waveguides, i.e., with a detector array aligned with the waveguides. 
Because we assume that the chosen base is orthonormal, the signal $I_n$ of the detector 
is
\begin{equation}
I_n = \left\lvert \sum_{i=1}^{N} c_i \psi_{n,i} \right\rvert^2,
\label{eq:refname2}
\end{equation}
which can be written in the following expanded form:

\begin{multline}
I_n = \sum_{i=1}^{N} |c_i|^2 |\psi_{n,i}|^2 \\+ 2 \sum_{i=1}^{N-1} \sum_{j=1}^{N} [\Re(c_i c_j^{*}) \Re(\psi_{n,i} \psi_{n,j}^{*}) - \Im(c_i c_j^{*}) \Im(\psi_{n,i} \psi_{n,j}^{*})],
\label{eq:refname3}
\end{multline}
where the operators $\Re(z)$ and $\Im(z)$ denote the real and imaginary parts of $z$, respectively. For all output waveguides, the equations in \eqref{eq:refname3} can be arranged in a matrix form:



\begin{equation}
\vec{I} =  \boldsymbol{V} \cdot\vec{J}.
\label{eq:refname4}
\end{equation}
By choosing the $N^2$-dimensional vector $\vec{J}$ as

\begin{multline}
\vec{J}=[|c_1|^2,\ldots,|c_N|^2,\\ \Re(c_1c_2^{*}),\ldots,\Re(c_{N-1}c_{N}^{*}),  \Im(c_1c_2^{*}),\ldots,\Im(c_{N-1}c_{N}^{*})]^T,
\label{eq:refname5}
\end{multline}
the matrix $\boldsymbol{V}$ can be divided into three sub-matrices $\boldsymbol{V} = [\boldsymbol{A}:\boldsymbol{B}:\boldsymbol{C}]$ where $\boldsymbol{A}$ is an $M \times N$ matrix that contains the squared moduli of $\psi_{n,i}$ while $\boldsymbol{B}$ and $\boldsymbol{C}$ are matrices of size $M \times N(N-1)/2$ that contain the real and imaginary parts of all possible pair products $\psi_{n,i}\psi^{*}_{n,j}$, respectively. The entries of $\boldsymbol{A}=[\alpha_{i,j}]$, $\boldsymbol{B}=[\beta_{i,j'}]$ and $\boldsymbol{C}=[\gamma_{i,j'}]$ are given explicitly by

\begin{align}
\alpha_{i,j}  &=  &|\psi_{i,j}|^{2},  \nonumber\\
\beta_{i,j'}  &=  &2\Re(\psi_{i,p(j')} \psi_{i,q(j')}^{*}),\nonumber\\
\gamma_{i,j'}  &=  &-2\Im(\psi_{i,p(j')} \psi_{i,q(j')}^{*}),\nonumber\\
&&&i =1, \cdots, M,\nonumber \\
&&&j = 1, \cdots, N,\nonumber \\
&&&j' = 1, \cdots, N(N-1)/2,
\label{eq:refname6}
\end{align}
where $p(j')$ and $q(j')$ are given by
\begin{multline}
p(j')  =  m, \quad \textnormal{ for } m \in \mathbb{N} \textnormal{ that satisfies}\\ (m-1)\left(N-\frac{m}{2}\right) < j' \leqslant m\left(N-\frac{m+1}{2}\right), \nonumber
\end{multline}
\begin{multline}
q(j') =  p(j') + j' +  [p(j')-1]\left[\frac{1}{2} p(j') - N\right] .
\label{eq:refname7}
\end{multline}

Now $\boldsymbol{V}$ in \eqref{eq:refname4} has a generalized inverse provided that its
number of rows $M\geq N^2$, i.e., the discrete beam combiner section of the device must at least have a number of coupled waveguides equal to the square of the number of modes to be analyzed.

While the numerical determination of the matrix $\boldsymbol{V}$ is straightforward, in a real 
experiment a calibration procedure is required. This can be accomplished with a method similar to 
the one used for the calibration of integrated optics interferometers for astronomical use (see 
for instance \cite{saviauk_3d-integrated_2013}). In the present case, a spatial light modulator is required to 
excite selectively the modes of the multimode end of the device and perform the calibration. In 
particular, the columns of the sub-matrix $\boldsymbol{A}$ are obtained from the power delivered by 
each output waveguide resulting from the excitation of individual modes of the photonic lantern.
The sub-matrices $\boldsymbol{B}$ and $\boldsymbol{C}$ are obtained by exciting the device with a 
superposition of modes with a linearly increasing phase delay between them. 


The column vector $\vec{J}$ [\eqref{eq:refname5}] contains the square moduli of the coefficients $c_i$ and the real and imaginary parts of all possible pair products of the modal amplitudes. Since only the relative phase between the modes matter to retrieve the input pattern of light, only the real and imaginary parts of the products of the amplitudes of a reference mode (\textit{e.g.} the fundamental one) by the conjugate of the amplitudes of the other modes are indeed necessary to solve the modal decomposition problem. Solving for $\vec{J}$ is equivalent to solving an overdetermined system of equations, which is possible by left multiplying the vector $\vec{I}$ by the Moore-Penrose pseudoinverse of $\boldsymbol{V}$: $\boldsymbol{V}^+=(\boldsymbol{V}^T\boldsymbol{V})^{-1}\boldsymbol{V}^T$ \cite{press_numerical_1992}. An important feature of this solution algorithm is to gauge the  sensitivity of the solution to the inevitable measurement noise of the elements of $\vec{I}$. This is possible by estimating the condition number of $\boldsymbol{V}$, which is defined as the ratio of the maximum to the minimum singular values of the matrix 
\cite{press_numerical_1992}. 

\label{subsec:optimization}

The condition number (which we denote with the symbol $\kappa$) can be interpreted as the ratio of the maximal to minimal stretching ratios of the matrix $\boldsymbol{V}$ along specific 
directions in the space \pagebreak of the $\vec{J}$ vector. Thus, an ill-conditioned matrix (large 
$\kappa$) would stretch the vectors much more in one direction than another meaning that 
small perturbations of the vector $\vec{J}$ chosen along certain directions would be 
amplified greatly by the mapping operation. Therefore, a small $\kappa$ is required 
to ensure the numerical stability of the transformation $\boldsymbol{V}$ and its pseudoinverse.
 
As it has been already shown in the past 
\cite{minardi_photonic_2012,minardi_nonlocality_2015}, the conditioning 
of the transfer matrix $\boldsymbol{V}$ of a discrete beam combiner significantly depends on the geometric 
parameters of the waveguide array and its excitation configuration.
To optimize the modal analyzer, we considered not only the geometry of the array of 
waveguides, but we also carried out tests to find a convenient injection geometry from the 
photonic lantern to the discrete beam combiner.
A thorough verification of all injection configurations would become very rapidly an 
intractable problem due to their factorial scaling with the number of waveguides in the 
photonic lattice. We therefore used a heuristic method to guide our optimization process, 
which was based on the requirement of a field transfer function $\psi_\mathrm{i,j}$ with aperiodic phases evenly distributed on the $0-2\pi$ interval \cite{minardi_nonlocality_2015}. 
The optimization was carried out by means 
of an RSoft CAD \cite{rsoft} model of the mode analyzer. The $\boldsymbol{V}$ matrices of the studied designs were constructed according to \eqref{eq:refname6} from the peak amplitudes of the fields at the waveguides centers $\psi_\mathrm{i,j}$ calculated by the beam propagation solver BeamPROP \cite{rsoft}. The multimode end of the device has an $8$ $\mu m$ core diameter and a numerical aperture of $0.17$ and hence supports $3$ LP modes (the fundamental LP$_{01}$ and the doubly-degenerate LP$_{11}$) at $\lambda = 1.5$ $\mu m$ (see Fig. \ref{fig:layout1}). 
The single-mode fibers emerging from the multimode core and the array's waveguides have a $5$ $\mu m$ core diameter each with a similar numerical aperture. The lattice constant of the hexagonal array is $7.5$ $\mu m$ which gives a coupling length of $L_C = 1$ $mm$. The lantern's taper angle is taken shallow enough for it to be adiabatic. 

\begin{figure}[b!]
\centering
\includegraphics[width=0.9\linewidth]{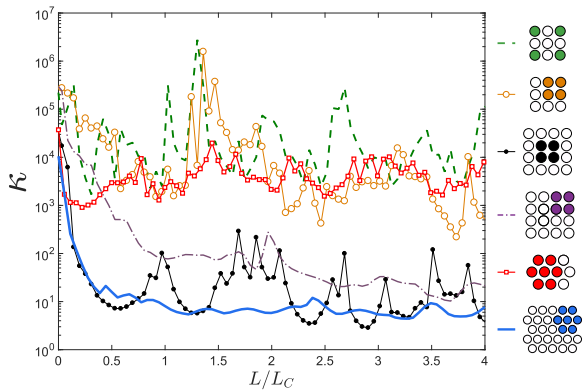}






\caption{(Color online) 
Variations of the condition number $\kappa$ of $\boldsymbol{V}$ along the array in units of coupling lengths for different geometries. A hexagonal oversized array with an off-center $1 \times 7$ lantern ( \protect\blue\space  curve) provides the best performance as it has a low condition number that is fairly insensitive to length.} 
\label{fig:CN_vs_L2}
\end{figure}

\begin{figure}[b!]
\centering
\includegraphics[width=0.9\linewidth]{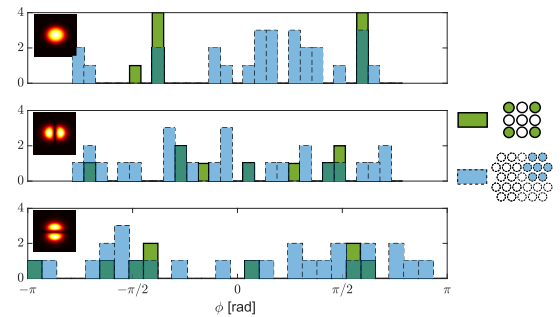}

\caption{(Color online) Histogram of the phase of the field transfer matrix $\psi_\mathrm{n,i}$ for i = 1, 2 and 3 (excitation of the device with the 3 modes supported by the lantern and shown in the insets) in a 9 waveguides square array 
(\space\protect\greenbox\space fill, $\kappa = 2 \times 10^6$) and a 27 waveguides hexagonal array with an off-center lantern (\space\protect\bluebox\space fill, $\kappa = 4.4$). Greater phase diversity of the field transfer matrix yields $\boldsymbol{V}$ matrices with lower condition numbers.}
\label{fig:histo2}
\end{figure}

Fig. \ref{fig:CN_vs_L2} shows the dependence of $\kappa$ on the array's length for some of the configurations we studied while attempting to optimize the device. This optimization strategy was supported by the observation that
an input configuration where the output waveguides of the photonic lantern are symmetrically 
distributed across the array give rise to highly ill-conditioned $\boldsymbol{V}$
matrices (one example of this case is the configuration shown by \green\space in Fig. \ref{fig:CN_vs_L2}). For those configurations the modes of the photonic lantern excite with high efficiency the analogous 
real-valued supermodes of the photonic lattice, which allow only two possible phases 
separated by $\pi$. 
On the contrary, by restricting the injection sites to contiguous sites of the photonic lattice (see \orange\space in Fig. \ref{fig:CN_vs_L2}), the field propagating in the array is in general a complex-valued superposition of 
different supermodes, allowing the transfer function $\psi_\mathrm{i,j}$ to take arbitrary phase values. An off-center excitation of the array provided a smoother dependence of $\kappa$ on the device's length 
(cf. \purple\space and \black\space in Fig. \ref{fig:CN_vs_L2})
as more phase diversity is introduced thanks to the broken symmetry of the system in agreement with previous 
results \cite{minardi_nonlocality_2015}.
Additionally, the array needed to be compatible with the symmetry of the modes of the photonic lantern to avoid modal losses \cite{birks_photonic_2015}. For this reason, we chose a $1\times 7$ photonic lantern supporting $N=3$ linearly polarized (LP) modes coupled to 7 neighboring sites of a hexagonal photonic lattice featuring more than $9$ waveguides (see \red\space in Fig. \ref{fig:CN_vs_L2}).
We thus significantly restricted the number of configurations to be tested with a beam propagation method to a bare minimum.
We confirmed that, even though the minimal number of waveguides required in the photonic lattice is $M = N^2$ (in this case, $9$ for the $3$ modes), an oversized device with more waveguides leads to better conditioned matrices $\boldsymbol{V}$ (cf. \red\space and \blue\space in Fig. \ref{fig:CN_vs_L2}). The phases at output waveguides corresponding to each of the supported modes for two of the configurations we studied are counted in the histograms in Fig. \ref{fig:histo2}. The oversized hexagonal configuration (shown by \blue\space in Fig. \ref{fig:CN_vs_L2} and by \bluebox\space in Fig. \ref{fig:histo2}) which possesses all the attributes we deemed beneficial to lowering $\kappa$ has its phases more evenly distributed in the interval $[-\pi,\pi]$ than the phases of the square configuration (shown by  \green\space in Fig. \ref{fig:CN_vs_L2} and by \greenbox\space in Fig. \ref{fig:histo2}), which lacks those attributes.   

Fig. \ref{fig:layout1} illustrates the layout of a device with an 
optimized configuration. The device has been designed for an input 
lantern supporting 3 modes and has a minimal condition number of 
$\kappa=4.4$ (at array length $L=3.3$ $mm$) as a result of the oversized array (27 
waveguides). Interestingly, the condition number varies smoothly 
along the length of the device (Fig. \ref{fig:CN_vs_L2},
\blue\space curve), so that the performance of a real device is relatively tolerant to uncertainties in the coupling coefficient of the waveguide array.\pagebreak

\begin{figure}[b!]
\centering
\includegraphics[width=0.88\linewidth]{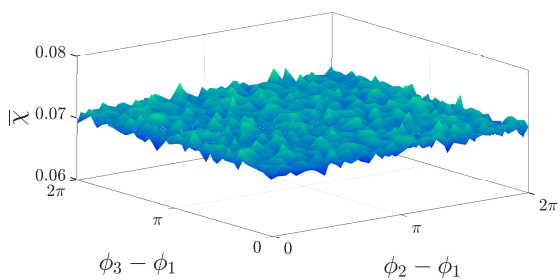}
\caption{The expected average relative error of the retrieved vector $\vec{J}$ from a noisy measurement of the output signal with the device depicted in Fig. \ref{fig:layout1}. The horizontal axes represent the relative phase between modes 1-2 and 1-3. Only an additive Gaussian noise with standard deviation $\epsilon=0.1$ was considered in the calculation (see text for details).}
\label{fig:chi2}
\end{figure}

The performance of the same device was evaluated numerically by calculating the impact of an additive random noise in the intensity measurement (\textit{e.g.} dark current of a single pixel detector) on the retrieval of the complex amplitudes of the input modes.  
The algorithm computes the intensity signal at the output of the mode analyzer $\vec{I}$ from an array of different input vectors $\vec{J}$, each one with unitary modal amplitude and a phase covering all possible combinations of relative phases between modes 1-2 and 1-3 sampled on the $[0, 2\pi]$ interval. We added to each element of the calculated $\vec{I}$ vectors a random number with zero mean and Gaussian distribution of amplitude $\epsilon$.
The noisy vector was transformed back to the modal amplitude vector $\vec{J}_\epsilon$ by means of the pseudoinverse of $\boldsymbol{V}$ and its average distance $\chi$ from the input vector $\vec{J}$ was calculated:

\begin{equation}
\chi = \sqrt{\frac{\sum_{i=1}^{N^2} [ J_{\epsilon}(i) - J(i) ]^2}{N^2}}.
\label{eq:refname8}
\end{equation}
The average of $\chi$ over 1000 noise realization was calculated for 
each input vector. Because of the chosen normalization, $\chi$ can 
be seen as the average relative error of the retrieved elements of vector $\vec{J}$.
Fig. \ref{fig:chi2} shows the averaged $\chi$ as a function of the
relative phases differences 1-2 and 1-3 for $\epsilon=0.1$. We 
notice that there is no preferential phase combination. The 
phase-averaged $\chi$ grows linearly with the noise $\epsilon$ with a slope of 0.7. 

The fidelity of the modal analysis depends ultimately on how well the modes of the multimode waveguide describe the input field. A numerical estimate shows that a lateral shift of the input field by 1/32 of the waveguide diameter will change the projected amplitudes of the modes by $\sim$10\%. 
We notice, however, that solid state lasers can feature a pointing stability better than 1/100 of the beam divergence. 

To conclude, we proposed a fast and computationally inexpensive method for determining 
the 
complex amplitudes of the excited modes in a waveguide by means 
of intensity measurements. A single matrix multiplication is required to retrieve the sought-after coefficients, which implies that the method can be as fast as the photo-detection is. 
The method is, in principle, applicable to any number of modes; however, the array must 
have a number of waveguides that is at least equal to the square of the number of modes. 
The geometry of the lantern's SMFs, the geometry of the array, the manner by which the lantern is 
connected to the array, and the array's length all have an impact on how the modes evolve 
into the supermodes of the array. This, in turn, determines the condition number of the 
device's transfer matrix, which must be low to have a system that is relatively 
insensitive to measurement noise. The optimal configuration, shown by \blue\space in 
Fig. \ref{fig:CN_vs_L2}, has a low $\kappa$ over a wide length range, which makes it 
suitable for broadband operation and makes it more tolerant to fabrication defects. 
A device like the one described here may be fabricated using 3D micro-fabrication techniques such as ultrafast laser inscription in glasses 
\cite{thomson_ultrafast_2009} or laser two-photon polymerization \cite{Woods:2014}. 

The device can be used to monitor the transverse modes in lasers for control purposes. The integrated mode analyzer could be matched directly to multimode 
fiber lasers or through free space optics to conventional laser cavities. The possibility to operate the device 
in real time 
opens its application in
the control of the beam quality of fiber lasers \cite{Otto:2013} or 
of the oscillating mode in transverse multi-stable lasers \cite{Tamm_bistability_1990}.\newline

\noindent
\textbf{Funding.} German Federal Ministry of Education and Research (BMBF) (03Z22AN11).

\bibliography{Main}

\bibliographyfullrefs{Main}
 

\ifthenelse{\equal{\journalref}{aop}}{%
\section*{Author Biographies}
\begingroup
\setlength\intextsep{0pt}
\begin{minipage}[t][6.3cm][t]{1.0\textwidth} 
  \begin{wrapfigure}{L}{0.25\textwidth}
    \includegraphics[width=0.25\textwidth]{john_smith.eps}
  \end{wrapfigure}
  \noindent
  {\bfseries John Smith} received his BSc (Mathematics) in 2000 from The University of Maryland. His research interests include lasers and optics.
\end{minipage}
\begin{minipage}{1.0\textwidth}
  \begin{wrapfigure}{L}{0.25\textwidth}
    \includegraphics[width=0.25\textwidth]{alice_smith.eps}
  \end{wrapfigure}
  \noindent
  {\bfseries Alice Smith} also received her BSc (Mathematics) in 2000 from The University of Maryland. Her research interests also include lasers and optics.
\end{minipage}
\endgroup
}{}

\end{document}